\documentclass[preprint,showpacs,preprintnumbers,amsmath,amssymb]{revtex4}
\usepackage{graphicx}
\usepackage{dcolumn}
\usepackage{bm}
\usepackage{CJK}
\usepackage{amsfonts}
\usepackage{psfrag}
\usepackage{wrapfig}
\usepackage{subfigure}
\usepackage{makeidx}
\usepackage{multirow}
\usepackage{epsf}
\usepackage{hyperref}
\usepackage{amsmath}

\usepackage{cases}

\begin{document}

\newcommand{\eq}[1]{Eq.(\ref{#1})}
\newcommand{\ud}{\,\mathrm{d}\,}

\title{Cosmological evolution of quintessence with a sign-changing interaction in dark sector}

\author{Ming Zhang$^1$, Cheng-Yi Sun$^2$\footnote{Corresponding author: ddscy@163.com}, Zhan-Ying Yang$^{1}$\footnote{zyyang@nwu.edu.cn},Rui-Hong Yue$^{3}$\footnote{yueruihong@nbu.edu.cn} }

\address{$^{1}$Department of Physics, Northwest University, Xi'an, 710069, China\\
$^{2}$Institute of Modern Physics, Northwest University, Xi'an 710069, China\\
$^{3}$Faculty of Science, Ningbo University, Ningbo 315211, China}

\begin{abstract}
  By analysing the cosmological dynamical system with a dark-sector interaction which changes sign during the cosmological evolution, we find a scaling attractor to help to alleviate the cosmic-coincidence problem. This result shows that this interaction can bring new features to the cosmology.

\end{abstract}


\maketitle

\section{Introduction}

Observations of Type Ia Supernova show us that the present expansion of our universe is being accelerated\cite{Perlmutter:1998np}. Typically the cosmological constant $\Lambda$ is used to explain the acceleration, but it suffers from the fine-tuning problem and the cosmic-coincidence problem\cite{Jassal:2005qc}. To alleviate the coincidence problem, Li and Zhang \cite{Li:2011ga} suggest the idea of scaling attractors by considering the cosmological dynamics without interaction between dark energy and dark matter. However, recently some observational evidences have suggested the possibility of the interaction between dark matter and dark energy.(\cite{Bertolami:2007zm}-\cite{Gumjudpai:2005ry}).

In order to consist with the facts, it is necessary to consider the cosmological dynamics with the interaction in the dark energy and dark matter. Motivated by the result in \cite{Cai:2009ht} that the interaction between dark energy and cold dark matter is likely to change the sign during the cosmological evolution, a new interaction between dark energy and dark matter has been suggested by Sun and Yue  \cite{Sun:2010vz}. This interaction is consistent with the second law of thermodynamics and the observational constraints, in that the signs of the interaction changes from negative to positive as the expansion of our universe changing from decelerated to accelerated. In this paper, we consider the cosmological evolution of quintessence with this interaction and find a scaling attractor to alleviate the cosmic-coincidence problem.

\section{Interacting Quintessence Energy With Dark Matter}

We consider a spatially flat FRW universe containing a scalar field $\phi$ and cold dark natter. The governing equations of motion are:
\begin{eqnarray}
  &&H^2\equiv (\frac{\dot{a}}{a})^2=\frac{\kappa^2}{3}(\rho_\phi+\rho_m),\label{equ:fde1}\\
  &&\dot{H}=-\frac{\kappa^2}{2}(\rho_\phi+p_\phi+\rho_m),\\
  &&\kappa^2\equiv 8\pi G
\end{eqnarray}
where $H$ is the Hubble parameter, $a(t)$ is the scale factor with cosmic time. A dot denotes the derivative with respect to the cosmic time $t$. We denote pressure and energy densities of the scalar field as $p_\phi$ and $\rho_\phi$ with an equation of state $\omega_\phi=p_\phi/\rho_\phi$. Likewise, $p_m$ and $\rho_m$ represent the pressure and energy densities of the dark matter with a similar equation of state $\omega_m=p_m/\rho_m$.

We postulate that the dark energy and dark matter interact through the interaction term $Q=3\sigma H(\rho_\phi-\rho_m)$ which we have mentioned above, so we have two conservation laws:
\begin{eqnarray}
  &&\dot{\rho}_m+3H\rho_m=Q\\
  &&\dot{\rho}_\phi+3H(\rho_\phi+p_\phi)=-Q
\end{eqnarray}
The parameter $\sigma$ is assumed to be positive.

Now we will consider the cosmological evolution of quintessence with this type of interaction. In this situation, we have:
\begin{eqnarray}
  \rho_d=\rho_\phi=\frac{1}{2}\dot{\phi}^2+V(\phi)\\
  p_d=p_\phi=\frac{1}{2}\dot{\phi}^2-V(\phi)\label{equ:pd}
\end{eqnarray}
$V(\phi)$ is the potential. In this work, we consider the exponential potential. The exponential potentials allow the possible existence of scaling solutions in which the scalar field energy density tracks that of the perfect fluid (so that at late times neither component can be negligible) \cite{Guo:2004fq}.
\begin{equation}
  V(\phi)=V_0e^{-\lambda \kappa \phi}
\end{equation}
We assume the dimensionless constant $\lambda$ is positive since we can make it positive through $\phi\rightarrow -\phi$ if $\lambda<0$.

\section{Dynamical System Of Interaction Quintessence}

We use the method of dynamical system to consider the cosmological evolution. Following others researchers \cite{Copeland:1997et,Wei:2010fz}, we define the following dimensionless variables:
\begin{eqnarray}
\label{equ:xyz}
  x\equiv \frac{\kappa\dot{\phi}}{\sqrt{6}H},~~
  y\equiv \frac{\kappa\sqrt{V}}{\sqrt{3}H},~~
  z\equiv \frac{\kappa\sqrt{\rho_m}}{\sqrt{3}H}
\end{eqnarray}

Then \eq{equ:fde1}-\eq{equ:pd} can be written in the following autonomous form:
\begin{eqnarray}
  &&\frac{\ud x}{\ud N}=(s-3)x+(\sqrt{\frac{3}{2}}\lambda y^2-Q_1),\\
  &&\frac{\ud y}{\ud N}=sy-\sqrt{\frac{3}{2}}\lambda xy,\\
  &&\frac{\ud z}{\ud N}=(s-\frac{3}{2})z+Q_2,
\end{eqnarray}
where $N=\ln{a}$ is the number of e-foldings which is convenient to use for the dynamics of dark energy,
and
\begin{eqnarray}
 s&=&3x^2+\frac{3}{2}z^2\\
 Q_1&=&\frac{\kappa Q}{\sqrt{6}H^2 \dot{\phi}}=\frac{3\sigma}{2x}-\frac{3\sigma z^2}{x}\\
 Q_2&=&\frac{z Q}{2H\rho_m}=\frac{3\sigma(x^2+y^2-z^2)}{2z}
\end{eqnarray}

 and the relation: $$x^2+y^2+z^2=1$$

\section{Solutions}

Using the definition of the fixed points
($\frac{\ud x_c}{\ud N}=0,~\frac{\ud y_c}{\ud N}=0$), we have:
\begin{subequations}
  \begin{numcases}{}
  (s-3)x_c+(\sqrt{\frac{3}{2}}\lambda y_c^2-Q_1)=0\\
  sy_c-\sqrt{\frac{3}{2}}\lambda x_cy_c=0
  \end{numcases}
\end{subequations}
We can obtain the solutions of the above equations:

1. $y_c=0$ then:
\begin{equation}
  x_c=\pm\sqrt{\frac{(2\sigma+1)\pm \sqrt{(2\sigma+1)^2-4\sigma}}{2}}
\end{equation}

2. $y_c\neq 0$ then we have one real solution of the cubic equation:

\begin{equation}
  x_5=\sqrt[3]{-\frac{q}{2}+\sqrt{(\frac{q}{2})^2+(\frac{p}{3})^3}}
     +\sqrt[3]{-\frac{q}{2}-\sqrt{(\frac{q}{2})^2+(\frac{p}{3})^3}}
     +\frac{\frac{2}{3}\lambda^2+4\sigma+2}{6\sqrt{\frac{2}{3}}\lambda}
\end{equation}
where
\begin{eqnarray}
  p=&-&\frac{(\frac{2}{3}\lambda^2+4\sigma+2)^2}{8\lambda^2}+\sigma+\frac{1}{2}\\
  q=&-&\frac{(\frac{2}{3}\lambda^2+4\sigma+2)^3}{144\sqrt{\frac{2}{3}}\lambda^3}
    +\frac{(\frac{2}{3}\lambda^2+4\sigma+2)(\sigma+\frac{1}{2})}{6\sqrt{\frac{2}{3}}\lambda}
    +\frac{\sigma}{2\sqrt{\frac{2}{3}}\lambda}
\end{eqnarray}
Note that these critical points must satisfy the conditions $y_c\geq 0$ and $z_c\geq 0$ by definitions in \eq{equ:xyz}. Thus we can derive the corresponding critical value of $y$:
\begin{equation}
  y_5=\sqrt{x_5^2-\sqrt{\frac{2}{3}\lambda x_5+1}}
\end{equation}
All of the five solutions are presented in Table I.

\begin{center}
\begin{table}[!hbp]
\caption{Five solutions of the dynamical system}
\begin{tabular}{|c|c|c|c|}
\hline
\hline
\scriptsize{label} & $x_c$ &  $y_c$ &  $z_c$  \\
\hline
\scriptsize{M1p} & $\sqrt{\frac{(2\sigma+1)+\sqrt{(2\sigma+1)^2-4\sigma}}{2}}$ & 0 & $\sqrt{\frac{(1-2\sigma)-\sqrt{(2\sigma+1)^2-4\sigma}}{2}}$\\
\hline
\scriptsize{M1m} & $\sqrt{\frac{(2\sigma+1)-\sqrt{(2\sigma+1)^2-4\sigma}}{2}}$ & 0 & $\sqrt{\frac{(1-2\sigma)+\sqrt{(2\sigma+1)^2-4\sigma}}{2}}$\\
\hline
\scriptsize{M2p} & $-\sqrt{\frac{(2\sigma+1)+\sqrt{(2\sigma+1)^2-4\sigma}}{2}}$ & 0 & $\sqrt{\frac{(1-2\sigma)-\sqrt{(2\sigma+1)^2-4\sigma}}{2}}$ \\
\hline
\scriptsize{M2m} & $-\sqrt{\frac{(2\sigma+1)-\sqrt{(2\sigma+1)^2-4\sigma}}{2}}$ & 0 & $\sqrt{\frac{(1-2\sigma)+\sqrt{(2\sigma+1)^2-4\sigma}}{2}}$\\
\hline
\scriptsize{M3} & $x_5$ & $y_5$ & $\sqrt{1-x^2-y^2}$ \\
\hline
\end{tabular}
\end{table}
\end{center}

\section{Stability Analysis}

Let $f(x,y)=\frac{\ud x}{\ud N},g(x,y)=\frac{\ud y}{\ud N}$, we can obtain the matrix
\begin{eqnarray}
  \mathcal{M}=\left(\begin{array}{cc}\frac{\partial f}{\partial x}&\frac{\partial f}{\partial y} \\ \frac{\partial g}{\partial x}&\frac{\partial g}{\partial y}\end{array}\right)_{(x=x_c,y=y_c)}
  =\left(\begin{array}{cc} A&B\\C&D \end{array}\right)\nonumber
\end{eqnarray}
The expressions of A,B,C,D are:
\begin{align}
  A&=\frac{9}{2}x_c^2-\frac{3}{2}y_c^2+\frac{3\sigma y_c^2}{x^2}
     -\frac{3\sigma}{2x_c^2}-3\sigma-\frac{3}{2},\nonumber\\
  B&=-3x_cy_c+\sqrt{6}\lambda y_c-\frac{6\sigma y_c}{x_c},\nonumber\\
  C&=3x_cy_c-\sqrt{\frac{3}{2}}\lambda y_c,\nonumber\\
  D&=\frac{3}{2}(x_c^2-y_c^2+1)-3y_c^2-\sqrt{\frac{3}{2}}\lambda x_c.\nonumber
\end{align}
The two eigenvalues of $\mathcal{M}$:
\begin{equation}
  \mu_{1,2}=\frac{\pm \sqrt{(A+D)^2-4(AD-BC)}}{2}+\frac{(A+D)}{2}\label{equ:mu}
\end{equation}
The stability around the fixed points depends on the nature of the eigenvalues:

1.Stable node: $\mu_1<0$ and $\mu_2<0$.

2.Unstable node: $\mu_1>0$ and $\mu_2>0$.

3.Saddle point: $\mu_1<0$ and $\mu_2>0$ (or $\mu_1>0$ and $\mu_2<0$).

In the case of $y_c=0$, we can get $B=C=0$. Putting this into the \eq{equ:mu}, we can derive $\mu_1=A,\ \mu_2=D \ $.
Thus we can give the eigenvalues of each solution. To make the results more intuitive, $r_1$ and $r_2$ are given by:
$$r_1=\frac{1}{2}(2\sigma+1), ~~ r_2=\frac{1}{2}\sqrt{(2\sigma+1)^2-4\sigma}$$.

\textbf{A.}~Point(M1p)
\begin{eqnarray}
  &&\mu_1=\frac{9}{2}(r_1+r_2)-(3r_1-\frac{3}{2})(\frac{1}{2(r_1+r_2)}+1)-\frac{3}{2}\\
  &&\mu_2=\frac{3}{2}(r_1+r_2+1)-\lambda\sqrt{\frac{3}{2}(r_1+r_2)}
\end{eqnarray}
It is unstable under the condition $\lambda<\sqrt{\frac{3}{2}}(x_c+\frac{1}{x_c})$, and saddle under the condition $\lambda>\sqrt{\frac{3}{2}}(x_c+\frac{1}{x_c})$ (because the second eigenvalue is negative).

\textbf{B.}~Point(M1m)
\begin{eqnarray}
  &&\mu_1=\frac{9}{2}(r_1-r_2)-(3r_1-\frac{3}{2})(\frac{1}{2(r_1-r_2)}+1)-\frac{3}{2}\\
  &&\mu_2=\frac{3}{2}(r_1-r_2+1)-\lambda\sqrt{\frac{3}{2}(r_1-r_2)}
\end{eqnarray}
It is stable under the condition $\lambda>\sqrt{\frac{3}{2}}(x_c+\frac{1}{x_c})$, and saddle under the condition $\lambda<\sqrt{\frac{3}{2}}(x_c+\frac{1}{x_c})$ (because the fist eigenvalue is negative).

\textbf{C.}~Point(M2p)
\begin{eqnarray}
  &&\mu_1=\frac{9}{2}(r_1+r_2)-(3r_1-\frac{3}{2})(\frac{1}{2(r_1+r_2)}+1)-\frac{3}{2}\\
  &&\mu_2=\frac{3}{2}(r_1+r_2+1)+\lambda\sqrt{\frac{3}{2}(r_1+r_2)}
\end{eqnarray}
We find that M2p is a saddle point for any $\lambda$ and $\sigma$ (the first eigenvalue is negative).

\textbf{D.}~Point(M2m)
\begin{eqnarray}
  &&\mu_1=\frac{9}{2}(r_1-r_2)-(3r_1-\frac{3}{2})(\frac{1}{2(r_1-r_2)}+1)-\frac{3}{2}\\
  &&\mu_2=\frac{3}{2}(r_1-r_2+1)+\lambda\sqrt{\frac{3}{2}(r_1-r_2)}
\end{eqnarray}
It is unstable for any $\lambda$ and $\sigma$.

Regarding the point (M3), which is beyond the scope of this paper, and it is the saddle point for any $\lambda$ and $\sigma$, which is not the solution we expect. The eigenvalues and stability of the five solutions are presented in the Table II.

\section{Conclusion}

It should be noted that Gabriela et al. \cite{CalderaCabral:2008bx} have studied a similar model in which the interaction parameters is taken to be both non-positive or both non-negative. However, the interaction parameters which we use in this paper have the opposite signs. With such an outcome, the scope of the model was generalized.

In summary, for the case with interaction $Q=3\sigma H(\rho_\phi-\rho_m)$, we find that there is one scaling attractor M1m which can help to alleviate the cosmological coincidence problem. The interactions herein suggested may bring new results to cosmology.

\begin{table}[!hbp]
\caption{The eigenvalues and stability of the five solutions}
\begin{tabular}{|c|c|c|c|}
\hline
\hline
label & $\mu_1$ & $\mu_2$ &  Stability  \\
\hline
M1p & A & D & $
\begin{array}{cll}
\text{Unstable node}&:&\lambda<\sqrt{\frac{3}{2}}(x_c+\frac{1}{x_c})\\
\text{Saddle point}&:&\lambda>\sqrt{\frac{3}{2}}(x_c+\frac{1}{x_c})
\end{array}$ \\
\hline
M1m & A & D & $
\begin{array}{cll}
\text{Saddle point}&:&\lambda<\sqrt{\frac{3}{2}}(x_c+\frac{1}{x_c})\\
\text{Stable node}&:&\lambda>\sqrt{\frac{3}{2}}(x_c+\frac{1}{x_c})
\end{array}$ \\
\hline
M2p & A & D & Saddle point:any $\lambda ,\sigma$ \\
\hline
M2m & A & D & Unstable node:any $\lambda ,\sigma$\\
\hline
M3 & $\mathcal{A}$ & $\mathcal{D}$ & Saddle point:any $\lambda ,\sigma$\\
\hline
\end{tabular}
$$\mathcal{A}=\frac{(A+D)+\sqrt{(A+D)^2-4(AD-BC)}}{2}$$
$$\mathcal{D}=\frac{(A+D)-\sqrt{(A+D)^2-4(AD-BC)}}{2}$$
\end{table}

\section{Acknowledgments}

This work has been supported by the National Natural Science Foundation of China (NSFC) (Grant Nos. 11147017 and 11347605),and the “Applied Nonlinear Science and Technology” of ZheJiang Province under Grant Nos. zx2012000070. and the ministry of education doctoral program funds (Grant No. 20126101110004).

\end{document}